\documentstyle[preprint,aps,epsf,rotate]{revtex}

\begin{document}
\input{epsf.sty}
\draft
\title{ Magnetic states induced by electron-electron interactions in a
plane quasiperiodic tiling}
\author{A. Jagannathan and H.J. Schulz}
\address{Laboratoire de Physique des
Solides, Universit\'{e} Paris--Sud, 91405 Orsay, France}
\date{September 1996}
\maketitle
%\widetext
%\vspace*{-1.truecm}
%\hspace*{-1.truecm}
\begin{abstract}
We consider the Hubbard model for electrons in a two-dimensional
quasiperiodic tiling using the Hartree--Fock approximation.  Numerical
solutions are obtained for the first three square approximants of the
perfect octagonal tiling. At half-filling, the magnetic state is
antiferromagnetic. We calculate the distributions of local
magnetizations and their dependence on the local environments as U is
varied.  The inflation symmetry of quasicrystals results in a corresponding
inflation symmetry of the magnetic configurations when passing from one
tiling to the next in the progression towards the infinite quasicrystal.
\end{abstract}
\hspace*{1.truecm}
%\pacs{PACS numbers: 71.27.+a}

\draft
%\Date{1/95}
%\debugon
%\narrowtext

Experimental studies of magnetism in quasicrystal-forming alloys
have shown that, for a given composition, the magnetic structure of
quasicrystalline, amorphous and crystalline phases were different. The
earliest studies, which were carried out on the intrinsically disordered
AlMnSi quasicrystals \cite{almn}, found that the quasicrystalline samples
had a small proportion of magnetic atoms and appeared to undergo a spin
glass transition. The crystalline phase of similar composition was
non-magnetic.  The early work stimulated interest in magnetic properties in
quasicrystalline media, because they indicated that moments were formed in
the quasicrystal whereas they were absent in close-lying crystalline phases.
The actual proportion of moment-carrying atoms was difficult to estimate in
the absence of any information as to the values of the moments and as to the
magnetic entities( possibly clusters of atoms).  Efforts were made
nevertheless to fit data obtained in NMR, M\"ossbauer and susceptibility
measurements \cite{prev}. The estimates in the case of AlMnSi as to the
proportion of the Mn atoms carrying moments vary from 1 \% to 60 \% .  The
underlying reason for these widely differing numbers is that the shape of
the distribution of moments is unknown, and is no doubt more complex than
the simple bimodal or Gaussian distributions implicitly assumed in analysis.
The early quasicrystals were relatively disordered, and could not be
improved by annealing, so that the role of disorder was difficult to
unentangle from that of quasiperiodicity. Since then, it has become possible
to study quasicrystals of far better structural quality, which are
thermodynamically stable, in contrast to their predecessors.  Recently,
results were reported for a ``good" quasicrystal AlPdMn \cite{hipp}. Moments
were observed to exist at high temperature, with a marked increase in value
at the transition to the liquid state. Again, the values and the
distribution of magnetic moments remain to be determined.

Motivated by this experimental situation, as well as the
intrinsic interest of a quasiperiodic Hubbard model, we have chosen to
investigate the influence of quasiperiodic order in a $single$ orbital
case. We consider the octagonal tiling, the ``fruit fly" of calculations on
quasiperiodic systems. It is a two-dimensional analogue of the 3D Penrose
tiling often encountered in models of real three-dimensional quasicrystals
and used to describe certain two-dimensional quasicrystals exhibiting
eight-fold symmetry in their diffraction spectrum.  The calculations were
carried out for the first three square approximants \cite{dun} of the
octagonal tiling, the perfect quasicrystal being obtained in the limit of
infinite size.  These square pieces are then periodically continued in the
plane. The tight-binding Hamiltonian matrix is given by
\begin{displaymath}
H = \sum_{i,j,s} 
 t_{ij} a^\dagger_{is} a^{\phantom{\dagger}}_{js}
+ U \sum_i n_{i\uparrow} n_{j\downarrow}
\end{displaymath}
where $a_{is}$ is the fermion destruction operator on site $i$ with spin
projection $s$. The energy scale is chosen so that $t_{ij}$ is one if
the sites are one tile length apart, and zero otherwise, and $U$ is the onsite
repulsion strength, multiplying the up- and down-spin densities. In the
Hartree--Fock approximation this second term is approximated by $U
\Sigma_\sigma \langle n_{i\sigma}\rangle n_{i-\sigma}$ , and the resulting
Hamiltonian matrix can then be diagonalized numerically. The iterative
method used consists of diagonalizing the Hartree--Fock Hamiltonian matrix
for an arbitrary initial distribution of electrons, finding the lowest
energy state, computing the Hartree--Fock interaction for the new
configuration and continuing until a self-consistent solution is attained.
As can be seen from our introduction of a quantization axis, rotational
symmetry has been explicitly broken, and a collinear structure is thus
assumed in this calculation. However, one can do a more general calculation
without this choice of parameterization --- indeed for fillings away from
the special values considered here, it is probably essential to allow for
non-collinear structures (see last section).

The model presented here is a precursor to models containing both s and d
orbitals, which would permit actual local moments as opposed to the
itinerant state of local magnetizations shown further below.  One could then
consider the situations where the $d$ orbitals are randomly placed (the
melt), or placed in well-defined sites, to investigate the corresponding
moments. Similar questions have been considered previously from an {\em ab
initio} standpoint in the melt,\cite{brat} for crystalline Hume--Rothery
alloys,\cite{tram} and for small clusters.\cite{liu} Those studies however
include the full complexity of the local atomic structure and therefore are
limited to rather small approximant phases. On the contrary, the relative
simplicity of the Hubbard model allows us to study bigger approximants and
therefore to concentrate on the effects of long--range quasicrystalline
order
on magnetism without being sidetracked by the effects of the local atomic
and electronic structure.

%The effect of local environment and short range order in the
%formation of moments in the melt has been recently investigated in numerical
%calculations on Al-Mn \cite{brat}, with the conclusion that the Mn atoms
%were all magnetic, contrary to the conclusions of the earlier cited studies
%on quasicrystals. {\it Ab initio} calculations \cite{tram} have considered
%the effect of $sp - d$ mixing in a number of crystalline Hume-Rothery alloys
%(the quasicrystal being considered a noncrystalline alloy of this
%type). These calculations confirmed that the moment formation is sensitively
%dependent on the details of the very complex electronic structure and is not
%simply related to the $d$-density of states of the transition metal ion. In
%another approach considering small systems, cluster calculations \cite{liu}
%have shown the strong dependence of the moment on the possible presence of a
%second Mn in the neighborhood.  This underlines the inadequacy of single
%impurity models. In an alloy, it is likely that an important role is played
%by frustration effects due to RKKY interactions between local moments in
%determining whether atoms are magnetic, and if so, the value of the magnetic
%moment.  In addition Kondo screening is an important issue in real
%quasicrystals. In \cite{hipp} the authors speculate that moments disappear
%in the quasiperiodic phase, due to the Kondo temperature being higher in the
%quasiperiodic phase compared to the liquid phase, where the spatial
%correlations are weakened. These issues remain of interest for future work.

In \cite{anu} the octagonal tiling was studied in the paramagnetic phase.
Within RPA, a magnetic transition was found to occur for a critical value
$U_c$ where $U_c$ depends on the Fermi energy. The values of the $U_c(E_F)$
fluctuate strongly, reflecting the huge fluctuations of the density of
states of the non-interacting model. More interesting was the observation
which are the sites ``most susceptible" to going magnetic. The answer
depends on the position of the Fermi level in the band, which is symmetric
for the bipartite case that we have studied here. For $E_F$ near the band
edges, the sites of high coordination number have the strongest divergence
of susceptibility at the transition.  (There exist six possible
coordination numbers ranging from 3 to 8.) For $E_F \approx 0$, i.e. near
the band center, on the other hand, it is the $low$ coordination sites
that are the most strongly susceptible at the $U$--driven magnetic
transition. Therefore it appears natural to expect that the new
magnetic state that we shall discuss in this paper will have big ``moments"
(hereafter we will use this term to denote the value of the local
magnetization density) associated with the high, or low coordination
numbers, depending on the value of the filling. This is indeed found, as
figures 1 and 5 illustrate below.

We now take up the case of half-filling: one electron per site, for tilings
of size 41, 239 and 1393 sites. (The actual sizes used in our calculations
are four times as large, as we have diagonalized $2 \times 2$ samples of
each approximant, to avoid distortions due to frustrating boundary
conditions).
Since each of the square approximants has in addition a
mirror plane, the number of non-equivalent sites is in fact smaller than the
numbers cited above. Concerning
 numerical difficulties of convergence to the correct
ground state: if one iterates using an arbitrary set of initial values of
the onsite moments, one finds several metastable solutions when $U$ is
 large. To overcome this problem, we have used the fact that the
sign of the moments is alternating on first neighbor sites, in making up
initial conditions that give rapid convergence to the ground state. For
arbitrary values of the filling, where the ground state is unknown this
method cannot be used.

At half-filling the nonmagnetic state obtained for $U=0$ gives way for
finite values of $U$ to an ``antiferromagnetic" state where moments on
adjacent sites (one hop distance apart) are anti-parallel.  This
antiferromagnetic ordering is to be expected because the tiling is bipartite
-- the sites and their links can be obtained by projecting a
four-dimensional cubic lattice down into the plane.  That the occupancy of
each site is exactly one (and equal on all sites - for half-filling only!)
is also easy to see. This is another consequence of the bipartite character
of the tiling, which ensures that not only the total density of states but
each $local$ onsite density of states is perfectly symmetric about the band
center (see plots in \cite{anu}).  For this case of one particle per site,
one can formulate the usual argument for minimizing the total energy in the
large $U$ limit, to obtain an effective antiferromagnetic Heisenberg model
with an energy of exchange proportional to $1/U$.

The onsite moments are distributed in a range of values. Figs.1a and 1b
illustrate the distribution of moments for the two bigger tilings. The radii
of the circles at each vertex of the tilings is proportional to the size of
the moment developed on that vertex, and the signs of the magnetization is
indicated by the color of the spots. Fig.1c shows the vertices of the
smallest approximant, from which the two successive tilings can be obtained
by one inflation/two inflations respectively. Note that the rings of moments
of Fig.1b are in one-to-one correspondence with the vertices of Fig.1c. This
is related to the fact that after two inflations $all$ vertices become the
centers of eight-fold rings.  Thus, the geometrical construction relating
successive approximants has its consequence for the magnetic structure,
which is also self-similar.  We note that the biggest circles, or largest
moments, correspond to sites of coordination 3 and 4, while the high
coordination sites have small values.  Fig.2 shows the distribution of
moments grouped according to the coordination number $z$. The smallest
moments correspond to the large $z$ sites, while the moments are largest for
the $z=3$ sites. Increasing the value of $U$ can be seen to shift the values
further towards the maximum possible values of $\pm 1$, while at the same
time reducing the width of the distribution of the values.  Fig. 2 also
shows an asymmetry between positive and negative moments, a feature more
easily seen in the detailed distribution of moments (below). This is due to
the finite size of the tilings, and one expects this asymmetry to be reduced
as tilings are increased in size, to reach the quasiperiodic limit.

In the limit of large $U$, the moments on all the sites tend to the value of 1
(which is the $exact$ occupancy of each site for $any$ value of $U$).  Fig.3
shows the approach to this asymptotic value for two different sites chosen
at random. The variation of each moment as a function of $U$ appears to be
site-dependent, and we have not found a compact way to characterize these
variations. Turning now to the details of the distribution of moments, we
have shown as an example, the histogram of moments for sites of $z=3$ in
Fig.4. The distribution shows a structure of peaks within peaks. A
distribution of values is of course to be expected, this being a reflection
of the fact that the local density of states is not identical for two sites
of the same coordination number, because of the differences in the
background. The peak structure arises because in a quasicrystal one can
always find a set of sites that have the same environment out to an
arbitrarily large distance. One sees that the distribution of moments
shrinks in width as $U$ gets larger, and the center of gravity shifts to
larger values of the moment. It is tempting to compare the histograms with
the plots in \cite{brat} for the case of Mn in a disordered background of
Al. In that work, the authors study the moments developed on Mn atoms in a
variety of environments in liquid Al-Mn. There, a ``bias" towards higher
values of the moment was remarked. This $may$ be related to the fact that
there are more sites of small coordination (for our octagonal tiling, as an
example, the density of $z=3$ sites is about 40\%, while that of $z=8$ sites
is about 3\%).

For small fillings, the Hartree--Fock ground state is ferromagnetic. The
occupancy ($n_\uparrow + n_\downarrow$) of sites is no longer uniform as for
half-filling but depends on the coordination number as well as on $U$. Fig.5
shows the second approximant (containing 239 sites), with circles of
diameter corresponding to the value of the moment at that site. A domain
wall separating the positive (filled circles) from negative (open circles),
appears, created by our imposition of a zero magnetization initial
condition.  The reflection-symmetry along the diagonal of our tiling imposes
the orientation of the domain wall.

Finally, we end with some remarks on what could be expected for other values
of the filling. We have found that convergence is very poor and trapping in
metastable states occurs.  One might expect the magnetic state to consist of
a mix negative and positive moments, or more likely, the moments may not
even be collinear. The numerical calculations are therefore substantially
bigger, as all spin orientations must now be allowed for. A new set of
calculations is presently underway to examine the magnetic state for
arbitrary fillings. One particular case that will be interesting to examine
is that of filling the band up to the pseudo-gap (at $E \approx \pm
2$). This may be closest to the experimental case, where a variety of
measurements as well as numerical calculations on ``realistic" models have
shown that $E_F$ lies in a local minimum of the density of states.  The
pseudogap at the Fermi energy observed (measured/calculated $E_F$ being a
third to a fifth of the expected free electron values \cite{ooo}) in
quasicrystalline and nearby crystalline alloys is thought\cite{frie} to be
due to a Hume-Rothery-Jones mechanism \cite{h-r}. The corresponding
band-filling for our octagonal tiling model is about 30\% (electrons or
holes). A study of the magnetic state for this case will be a necessary
further step in the understanding of the magnetism of quasicrystalline
magnetism.

{\bf Acknowledgments} We acknowledge gratefully the computing time and
facilities provided by IDRIS, Orsay

\newpage
{\centerline{ Figure Captions}}
\vskip 1cm
Fig. 1a) 239 sites approximant tiling showing moments at half-filling.
Circle radii correspond to moment size, filled/empty circles correspond to
positive/negative moments. $U=0.5$.

Fig. 1b) 1393 site tiling (links between sites are not shown) for
half-filling. Note the similarity with Fig. 1a.

Fig. 1c) The first tiling of the series, from which the other two can be
deduced. Note the position of the rings of moments of Fig. 1a coinciding
with the vertices of this tiling.

Fig. 2) Values of the onsite moments (positive and negative) plotted against
the site coordination number for two different $U$.

Fig. 3) Plot of moment versus $U$ for a selection of sites (arbitrary)

Fig. 4) Histogram of moment values at given values of $U$ for the sites of
coordination number z=3

Fig. 5) Magnetizations on the 239 site tiling showing the ferromagnetic
state for small filling (4 electrons)

\newpage {\large Fig.1a}
\centerline{
\epsfxsize=13cm
\rotate[r]{\epsffile{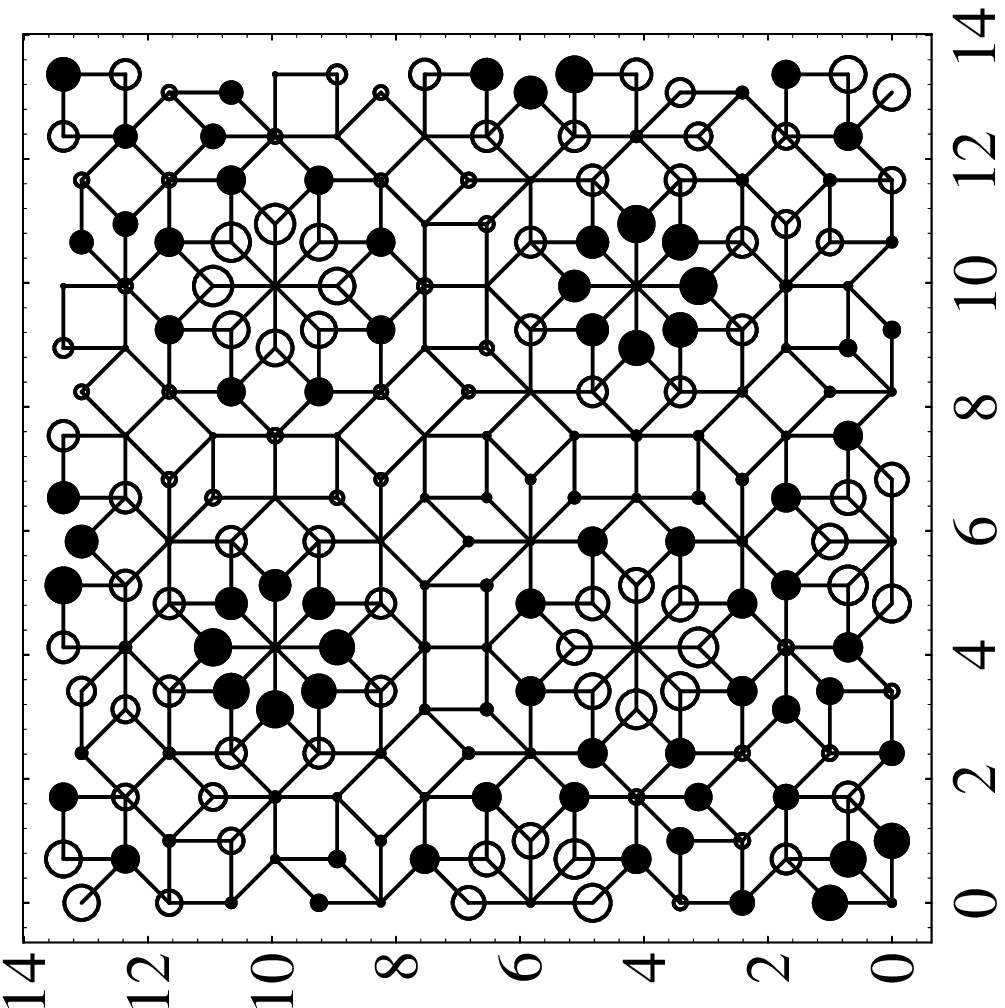}}
}
\newpage
{\large Fig. 1b}
\centerline{
\epsfysize=13cm
\rotate[r]{\epsffile{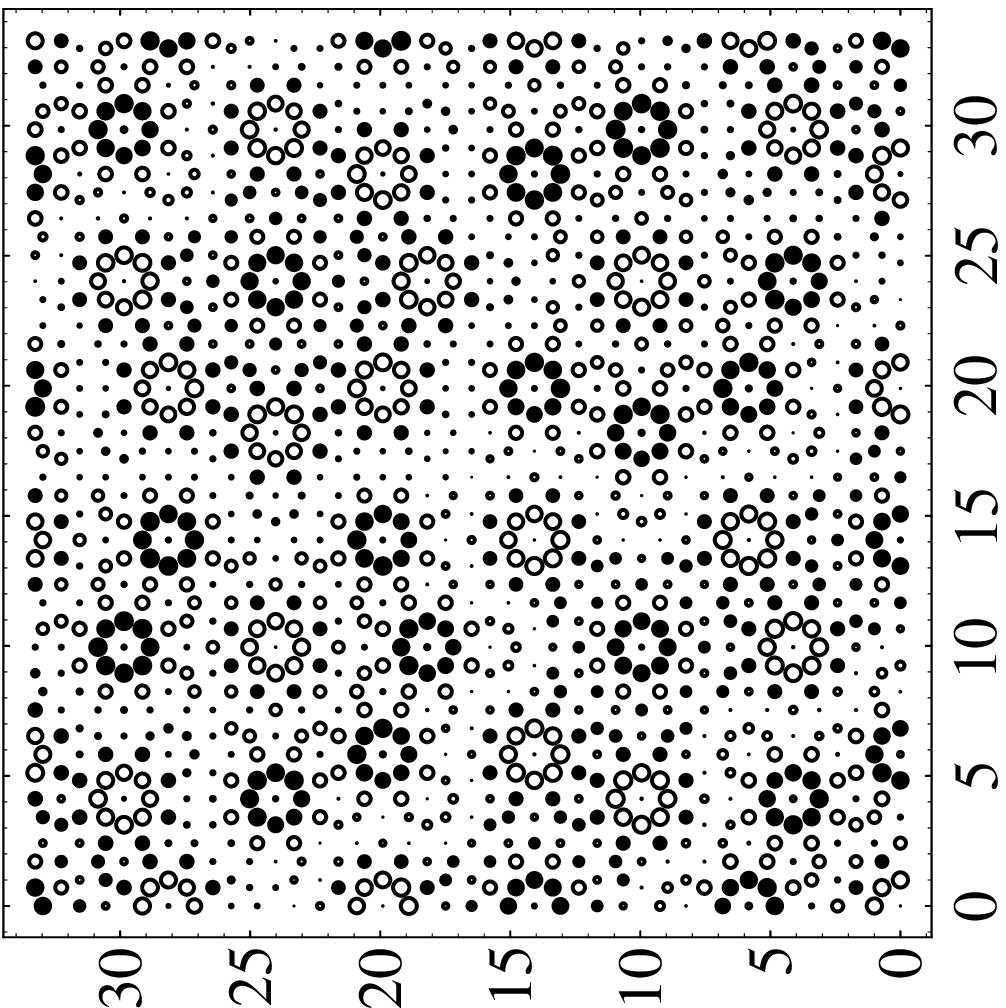}}}
{\large Fig. 1c}
\centerline{
\epsfysize=13cm
\rotate[r]{\epsffile{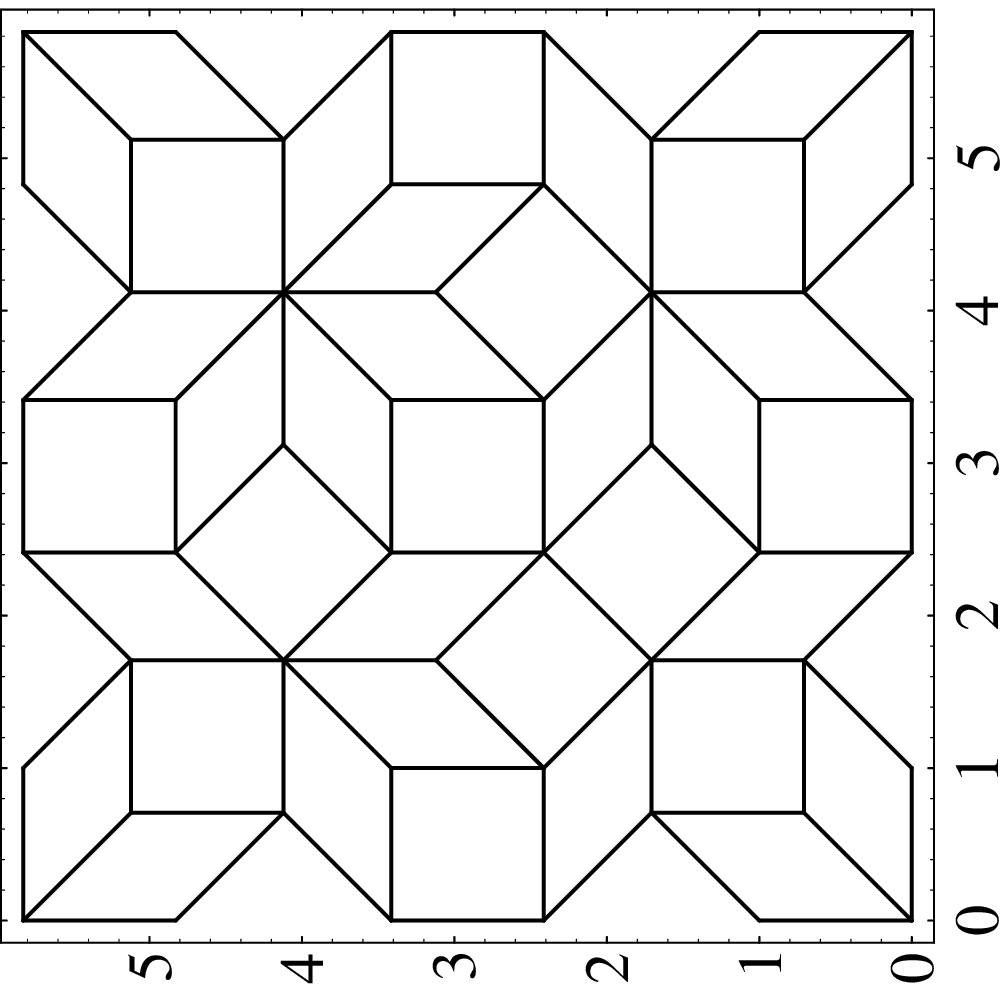}}
}
\newpage
{\large Fig. 2}
\centerline{
\epsfysize=20cm
\rotate[r]{\epsffile{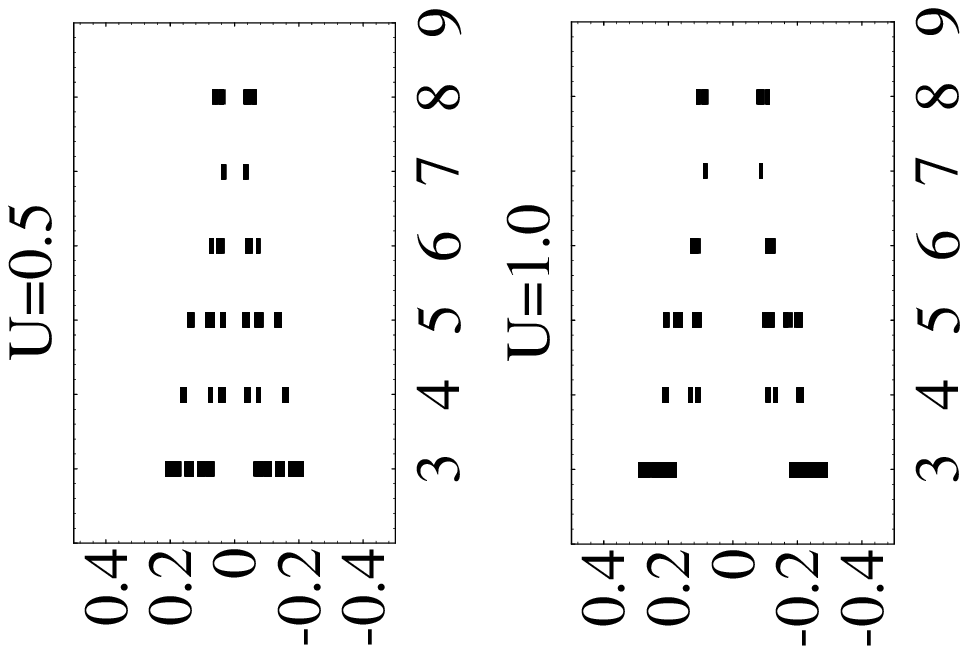}}
}
\newpage
{\large Fig. 3}
\centerline{
\epsfysize=15cm
\rotate[r]{\epsffile{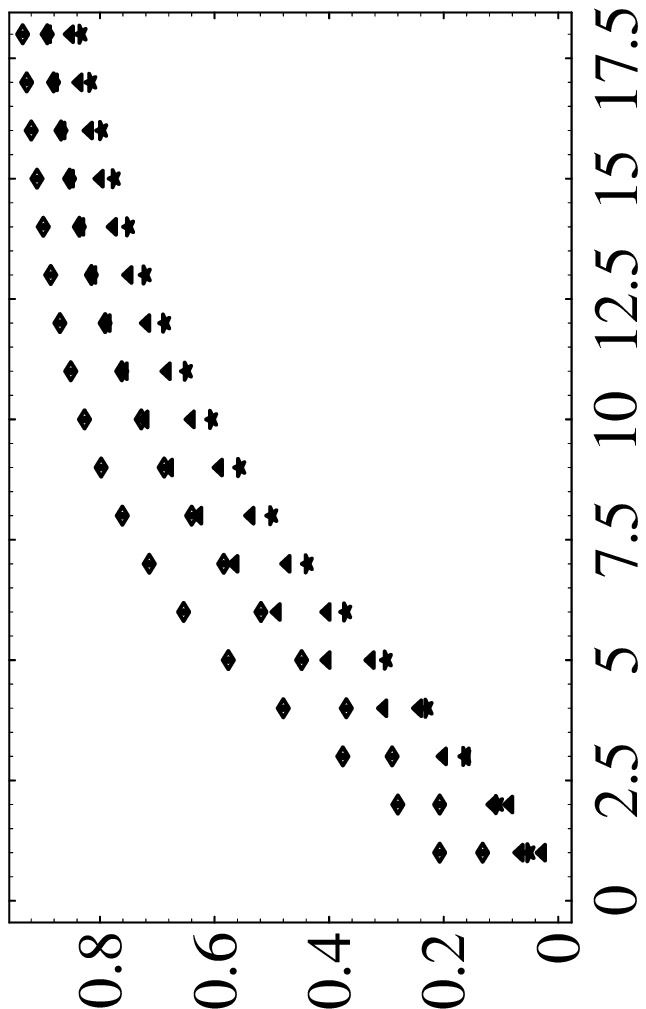}}
}
\newpage
{\large Fig. 4}
\centerline{
\epsfysize=15cm
\rotate[r]{\epsffile{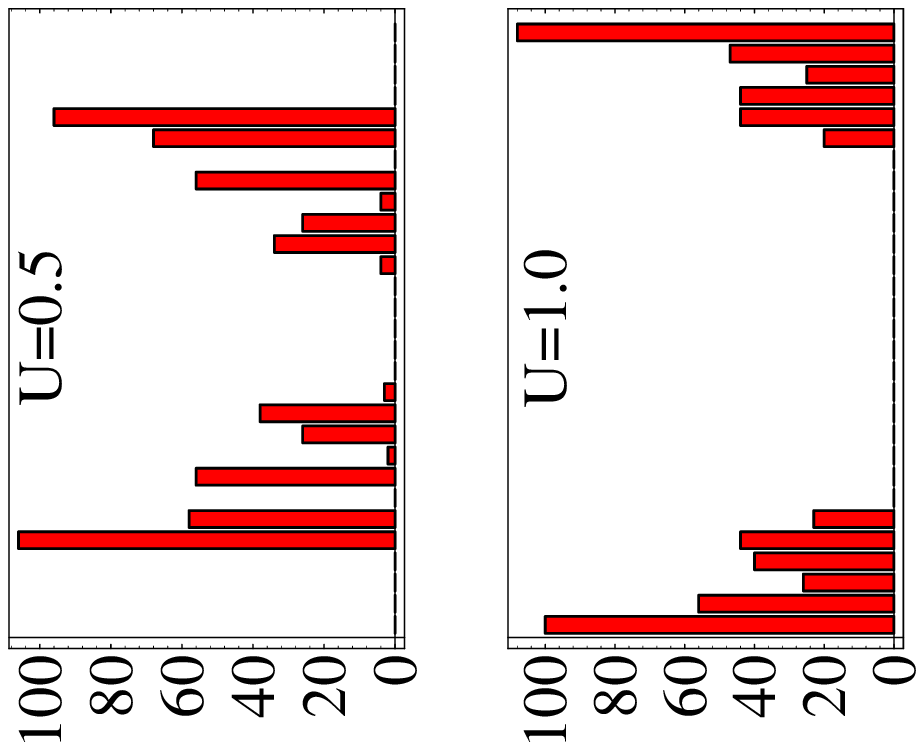}}
}
\newpage
{\large Fig. 5}
\centerline{
\epsfysize=15cm
\rotate[r]{\epsffile{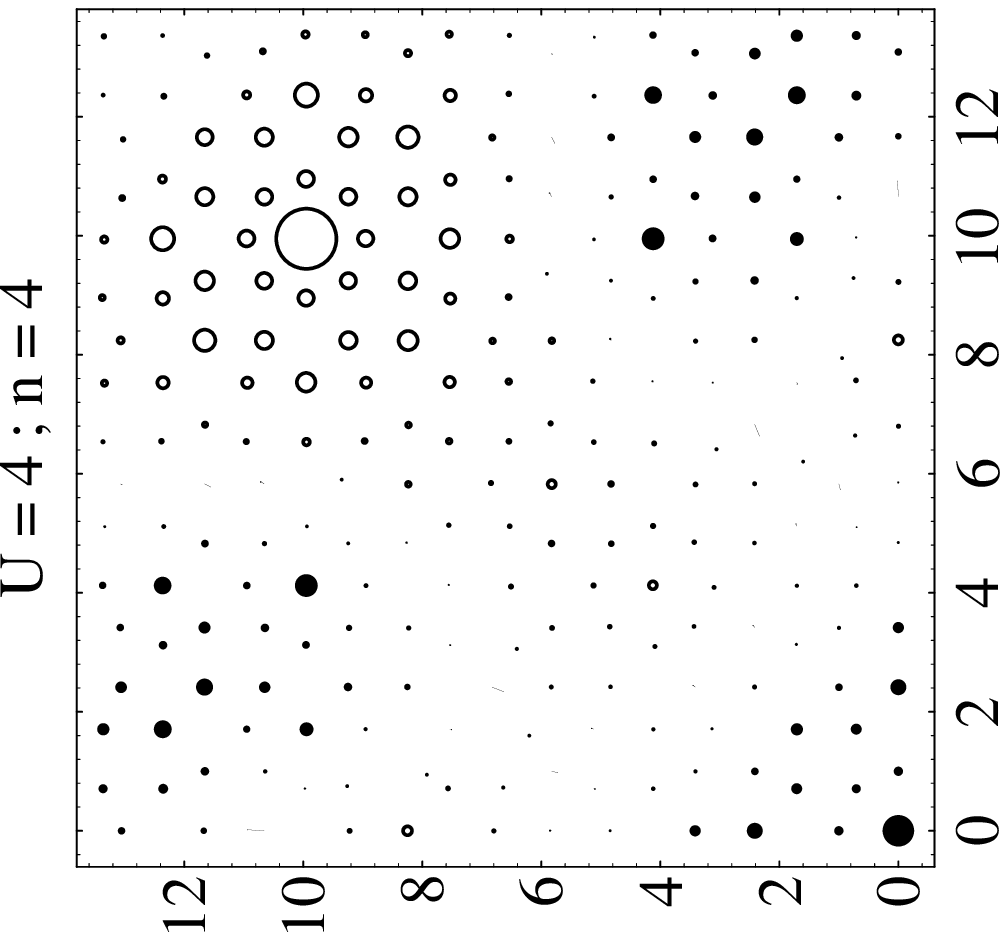}}
}
\end{document}